\documentclass[a4paper,10pt]{article}
\usepackage{amsmath,amssymb,epsfig,graphics}
\usepackage{caption}
\DeclareMathOperator{\sech}{sech}

\def\be{\begin{equation}}
\def\ee{\end{equation}}

\def\lb{\label}
\def\e{{\rm e}}
\def\ce{{\cal E}}
\def\cb{{\cal B}}

\def\lb{\label}

\def\e{{\rm e}}

\def\Sm{\Sigma_-}

\def\Nc{N_\times}
\def\Sp{\Sigma_+}

\def\Smh{{\Sigma_-^H}}

\def\Sph{{\Sigma_+^H}}

\usepackage{authblk}
\title{The electromagnetic spike solutions}
\author[1,2,4]{Ernesto Nungesser  \footnote{ernesto@maths.tcd.ie}}
\author[3,4]{Woei Chet Lim \footnote{wclim@waikato.ac.nz}}
\affil[1]{School of Mathematics, Trinity College, Dublin 2, Ireland}
\affil[2]{Department of Mathematics, Royal Institute of Technology, 10044~Stockholm, Sweden}
\affil[3]{Department of Mathematics, University of Waikato,
 Private~Bag~3105, Hamilton, New Zealand }
\affil[4]{Max-Planck-Institute for Gravitational Physics, Am~M\"{u}hlenberg~1, 14476 Golm, Germany}

\begin{document}

\maketitle

\begin{abstract}
The aim of this paper is to use the existing relation between polarized electromagnetic Gowdy spacetimes and vacuum Gowdy spacetimes to find explicit solutions for electromagnetic spikes by a procedure which has been developed by 
one of the authors for gravitational spikes. We present new inhomogeneous solutions which we call the EME and MEM electromagnetic spike solutions.
\end{abstract}

\section{Introduction}
According to Belinskii, Khalatnikov and Lifshitz (BKL)  \cite{BKL3,BKL1,BKL2}, a generic spacelike singularity is characterized
by asymptotic locality. Asymptotically towards the initial singularity each spatial point evolves independently
from its neighbors in an oscillatory manner that is represented by a sequence of Bianchi type I and II
vacuum models. In \cite{BergerMoncrief} Berger and Moncrief studied $T^3$-Gowdy spacetimes numerically and
observed the development of large spatial derivatives near the singularity, which they called `spiky
features'. These structures where found to occur in the neighborhood of isolated spatial surfaces, cf. \cite{GM}. Further numerical investigations (see \cite{Bergerliving} for an overview) seemed to 
indicate that the BKL conjecture is correct generically, but certain difficulties arose in simulating these spikes. An important step was made in \cite{RX} 
where a solution generating technique and Fuchsian methods developed in \cite{KR,RD} where used to produce asymptotic expansion for spikes, which where classified in `true' and
`false' spikes, where the latter are only a rotation artifact. Based on these transformations, in \cite{Lim} an explicit spike solution was found in terms of elementary
functions. The explicit spike solution suggests a new way to simulate spikes numerically, and in \cite{Lim2} it was confirmed that the explicit spike solution indeed describes the spiky structures, and does so remarkably 
accurately. The numerical results provide very strong evidence that apart from local BKL behavior, there also exist formation of spatial structures at and in the neighborhood of certain spatial surfaces, thus breaking asymptotic 
locality. 
Moreover the complete description of a generic spacelike singularity should involve spike oscillations \cite{HUL}, which are described by sequences of spike solutions and rotated Kasner solutions. 

We are interested in investigating the nature of BKL behaviour at spacetime singularities 
in the presence of electromagnetic fields. The aim of this paper is to use the known 
relation between polarized/diagonal electromagnetic Gowdy spacetimes
and Gowdy spacetimes, as given in \cite{ER}, to find spike solutions for the electromagnetic case. According to the BKL-picture a generic spacelike singularity is vacuum dominated.
This project thus will help to clarify whether the introduction of a Maxwell field changes the picture or not. 

The sign conventions of \cite{RA} are used. In particular, we use metric signature $- + + +$ and geometrized units.
We use the sign convention $\epsilon^{0123} > 0$ for the Levi-Civita tensor, but because we will switch to using a time variable that increases towards the past, 
the component $\epsilon^{0123}$ with respect to that coordinate system will be negative. 

\section{Basic equations and metric}

If $(\bar{P},\bar{Q},\bar{\lambda})$ is a solution of the vacuum Einstein equations with Gowdy symmetry (or orthogonally transitive $G_2$ isometry), i.e. with line element
\be
 ds^2=-\e^{(\bar{\lambda}-3\tau)/2}(d\tau^2+\e^{2\tau}dx^2)+\e^{-\tau}[\e^{\bar{P}}(dy+\bar{Q}dz)^2+e^{\bar{P}}dz^2],
\ee
 then $(P,\chi,\lambda)$ with (9)--(10) of \cite{ER} [with the following correspondence $t=e^{-\tau}$ and $(\theta,x,y)=(x,y,z)$]:
 \be
\label{NR_transform}
	P=2\bar{P}-\tau\,\quad
	\lambda=4\bar{\lambda}+4\bar{P}-\tau,\quad
	\chi = \bar{Q},
\ee
will be a solution of the Einstein-Maxwell equations with polarized Gowdy symmetry (or diagonal $G_2$ isometry) with line element:  
\be  \label{metric}
 ds^2=\e^{(\lambda-3\tau)/2}(d\tau^2+\e^{2\tau}dx^2)+\e^{-\tau}(\e^{P}dy^2+e^{-P})dz^2,
\ee
and a Maxwell field described by the vector potential with only one non-zero component, namely $A_3=\chi(x,\tau)$. We will assume from now on that we are in the second case, i.e. our metric is described via (\ref{metric}). The Einstein-Maxwell equations with $P$ and $\chi$ are given by (19)--(24) of \cite{nungesser}:
\begin{eqnarray*}
 &&P_{\tau\tau}-\e^{-2\tau}P_{xx}=2(\chi_{\tau}^2\e^{P+\tau}-\chi_x^2\e^{P-\tau})\\
&&\chi_{\tau\tau}-\e^{-2\tau}\chi_{xx}=\e^{-2\tau}P_x\chi_x-(P_{\tau}+1)\chi_{\tau}\\
&&\lambda_{\tau}=-P_{\tau}^2-\e^{-2\tau}P_x^2-4(\chi_{\tau}^2\e^{P+\tau}+\chi_x^2e^{P-\tau})\\
&&\lambda_x=-2P_{\tau}P_x-8\e^{P+\tau}\chi_{\tau}\chi_x.
\end{eqnarray*}
We will use the non-vanishing $\beta$-normalized variables \cite{VUW}.
\be
	\beta=\frac12 \e^{-\frac{\lambda-3\tau}{4}}.
\ee
We take this opportunity to clarify the sign confusion in Eq. (9) of \cite{Lim}. $\beta$ and other kinematic variables are defined with respect to the future-pointing congruence. Therefore
a positive $\beta$ describes expansion towards the future (and contraction towards the past). We also take this opportunity to correct the error in the $\beta$ expression in Eq. (2) of \cite{Lim2}.

These $\beta$-normalized variables refer to the $\beta$-normalized commutation functions associated with an orthonormal frame:
\[
	\Sigma_{\alpha \beta} = \frac{\sigma_{\alpha\beta}}{\beta},\quad
	N_{\alpha \beta} = \frac{n_{\alpha\beta}}{\beta}.
\]
The electric and magnetic fields are similarly normalized:
\[
	\ce_\alpha = \frac{E^\alpha}{\beta},\quad
	\cb_\alpha = \frac{B^\beta}{\beta}.
\]
The 3-by-3 $\Sigma_{\alpha\beta}$ and $N_{\alpha\beta}$ matrices in our case are
\begin{align}
\lb{matrix}
        \Sigma_{\alpha \beta}
        &= \left(\begin{matrix}
        -2\Sp & 0 & 0 \\
        0 & \Sp + \sqrt{3} \Sm & 0 \\
        0 & 0 & \Sp - \sqrt{3} \Sm
        \end{matrix}\right)
\\
\lb{matrix2}
        N_{\alpha \beta}
        &= \left(\begin{matrix}
        0 & 0 & 0 \\
        0 & 0 & \sqrt{3} \Nc \\
        0 & \sqrt{3} \Nc & 0
        \end{matrix}\right),
\end{align}
while the non-zero electric and magnetic components are
\[
	\ce = \frac{E^3}{\beta},\quad
	\cb = \frac{B^2}{\beta}.
\]
The $\beta$-normalized variables $\mathbf{Y}=(\Sigma_+,\Sigma_-,N_{\times},\ce,\cb)$ are related to the partial derivatives of $P$, $\lambda$ and $\chi$ as follows.
\be
\label{relation}
\mathbf{Y}=(\frac12+\frac16 \lambda_{\tau},-\frac{P_{\tau}}{\sqrt{3}},-\frac{\e^{-\tau}P_x}{\sqrt{3}},2\chi_{\tau}e^{\frac12(P+\tau)},-2\chi_xe^{\frac12(P-\tau)})
\ee

The evolution equations for the $\beta$-normalized variables can be derived from the above Einstein-Maxwell equations and (\ref{relation}):
\begin{eqnarray*}
&&\partial_t \Sigma_-=\e^{-\tau}\partial_x(N_{\times})+\frac{1}{2\sqrt{3}}(\cb^2-\ce^2),\\
&&\partial_t N_{\times}=\e^{-\tau}\partial_x \Sigma_- - N_\times,\\
&&\partial_t \ce=-\e^{-\tau}\partial_x \cb+\frac12\sqrt{3}N_{\times}\cb+\frac12(\sqrt{3}\Sigma_--1)\ce,\\
&&\partial_t \cb=-\e^{-\tau}\partial_x \ce-\frac12\sqrt{3}N_{\times}\ce-\frac12(\sqrt{3}\Sigma_-+1)\cb.
\end{eqnarray*}
We do not use the evolution equation for $\Sigma_+$, instead using the Gauss constraint to find $\Sigma_+$:
\begin{eqnarray*}
 \Sigma_+=\frac{1}{2}[1-\Sigma_-^2-N_{\times}^2-\frac{1}{3}(\ce^2 +\cb^2)].
\end{eqnarray*}

\section{Explicit solutions}

We now apply the vacuum-to-electromagnetic transformation (\ref{NR_transform}) to the explicit solutions in Section 4 of \cite{Lim}.

\subsection{Homogeneous solutions}
\subsubsection{Reparameterized Kasner solution}

Applying the transformation (\ref{NR_transform}) to the Kasner seed solution (17) of \cite{Lim} yields
\be
\label{Kasner_reparam}
	P=v\tau+2P_0,\quad
	\chi=\chi_0,\quad
	\lambda=-v^2\tau+4(\lambda_0+P_0),
\ee
where $v=2w-1$ and $P_0$, $\chi_0$ and $\lambda_0$ are arbitrary constants.
The result is trivial -- this solution is just a re-parametrization of the Kasner solution.
Nevertheless, we will use this parametrization of the Kasner solution in Figure~\ref{fig:EM_family} later.

The $\beta$-normalized variables $\mathbf{Y_K}$ have the same form as (18) of \cite{Lim}:
\be
 \mathbf{Y_K}=(\frac12 -\frac16 v^2,-\frac{v}{\sqrt{3}},0,0,0).
\ee

\subsubsection{Electric Rosen solution}
Applying (\ref{NR_transform}) to the rotated Kasner solution (22) of \cite{Lim} with the simplifying choice $P_0+\ln \chi_0=0$ yields
\begin{eqnarray}
\label{ERsol_1}
&&P=-\tau-2\ln\sech w \tau+2\ln 2 \chi_0\\
\label{ERsol_2}
&&\chi=-\frac{1}{2\chi_0}(1+\tanh w \tau)\\
&&\lambda=-(4w^2+1)\tau -4 \ln \sech w \tau +4(\lambda_0+\ln2\chi_0).
\label{ERsol_3}
\end{eqnarray}
In terms of the $\beta$-normalized variables,
\be
 \mathbf{Y_{ER}}=(\frac13+\frac23w(\tanh w \tau-w),\frac{1}{\sqrt{3}}(1-2w \tanh w \tau),0,-2w \sech w\tau,0).
\ee
This describes an electric Bianchi I model. We refer to \cite{CA} for a presentation of this solution in a broader context. Up to the arbitrary constant and choosing $\chi_0=1$, $\xi=e^{-\tau}$
we have exactly the same $\chi$ as described in \cite{CA} with a minus sign where now  $w=\frac{\alpha_0}{2}$ and $\gamma_0=\lambda_0$. This solution was first found by Gerald Rosen \cite{Rosen62} and we will refer to it as the electric Rosen solution with subscript $\text{ER}$.

\subsubsection{Magnetic Rosen solution}
Applying (\ref{NR_transform}) to the Taub solution (24)--(26) of \cite{Lim} yields
\begin{eqnarray}
\label{MRsol_1}
 &&P=\tau+2\ln\sech w\tau-2\ln 2 \chi_0\\
\label{MRsol_2}
&&\chi=2\chi_0 w x+\chi_1\\
&&\lambda=-(4w^2+1)\tau -4 \ln \sech w \tau +4(\lambda_1-\ln 2 \chi_0).
\label{MRsol_3}
\end{eqnarray}
We assume now that $\chi_1=0$ for simplicity.
In terms of the $\beta$-normalized variables,
\be \label{magneticrosen}
 \mathbf{Y_{MR}}=(\frac13+\frac23w(\tanh w \tau-w),\frac{1}{\sqrt{3}}(2w \tanh w \tau-1),0,0,-2w \sech w\tau).
\ee
We will refer to it as the magnetic Rosen solution and use the subscript $\text{MR}$. In particular 
we will use in the following the variables $(\Sigma_-)_\text{MR}=\frac{1}{\sqrt{3}}(2w \tanh w \tau-1)$ and $\cb_\text{MR}=-2w \sech w\tau$. The solution (\ref{magneticrosen}) is sometimes called the pure magnetic or magnetovacuum solution since later it has been generalized to include, e.g. a perfect fluid. 
It is not surprising that this solution appears here instead of the Bianchi II vacuum solution since there is a natural correspondence between heteroclinic chains consisting of Bianchi type II solutions in the vacuum case and 
heteroclinic chains in the case with a magnetic field which include orbits corresponding to both solutions of the vacuum Einstein equations of Bianchi type II and solutions of the Einstein-Maxwell equations of Bianchi type I. 
For recent work on oscillatory singularities in Bianchi models with magnetic fields we refer to \cite{Li2}. 

\subsection{Inhomogeneous solutions}
We will now proceed to the derivation of the inhomogeneous solutions. It is convenient to introduce the following definitions:
\begin{eqnarray}\label{spikyvar}
 f=x w \e^{\tau}\sech w \tau, \ \  s=\frac{2f}{f^2+1}, \ \ c=\frac{f^2-1}{f^2+1},
\end{eqnarray}
where it holds that $c^2+s^2=1$.

\subsubsection{EME electromagnetic spike solution}
Applying (\ref{NR_transform}) to the rotated Taub solution (28)--(30) of \cite{Lim}
\footnote{As pointed out on page 14 of \cite{Lim2}, the third minus sign in equation (28) of \cite{Lim} should be a plus sign, and the factor 4 in equation (34) of \cite{Lim} 
should not be there.} we obtain a new, inhomogeneous solution:
\begin{eqnarray}
\label{EME__1}
&&P=-3\tau-2\ln\sech w\tau+2\ln(f^2+1)+ 2\ln 2 \chi_0\\
\label{EME__2}
&&\chi=-\frac{1}{2\chi_0}\frac{f^2}{(f^2+1)x w}\\
\label{EME__3}
&&\lambda=-(4w^2+9)\tau -12 \ln \sech w \tau +4\ln(f^2+1)+4(\lambda_1+\ln2 \chi_0).
\end{eqnarray}
The spike occurs at $x=0$, and the electric field is zero there.
We will refer to the solution as the EME spike solution, because (for $|w| >1$ cases) worldlines along large $x$ experience a sequence of three Rosen transitions: electric-magnetic-electric.
Its $\beta$-normalized variables are given by
\be
 \mathbf{Y_{EME}}=[-\frac13(1+2w^2-2c+\sqrt{3}(c-2)\bar{\Sigma}_-),\frac{1}{\sqrt{3}}+c \bar{\Sigma}_-,s\frac{\cb_\text{MR}}{\sqrt{3}},s\sqrt{3}\bar{\Sigma}_-,c \cb_\text{MR}]
\ee
where we have denoted $\bar{\Sigma}_-=(\Sigma_-)_\text{MR}-\frac{1}{\sqrt{3}}$, and the subscripts $\text{MR}$ and $\text{EME}$ refer to the magnetic Rosen and the EME spike solution.

\subsubsection{MEM electromagnetic spike solution}
Applying (\ref{NR_transform}) to the spike solution (33)--(35) of \cite{Lim} yields
\begin{eqnarray}
\label{MEM__1}
 &&P=3\tau+2 \ln \sech w \tau -2 \ln (f^2+1)-2\ln 2 \chi_0\\
\label{MEM__2}   
&&\chi=-\chi_0 w [\e^{-2\tau}+2(w \tanh w \tau -1)x^2]+\chi_2\\
\label{MEM__3}
&&\lambda=-(4w^2+9)\tau -12 \ln \sech w \tau +4\ln(f^2+1)+4(\lambda_2-\ln2 \chi_0).
\end{eqnarray}
We will refer to it as the MEM spike solution and use the subscript $\text{MEM}$.
Its $\beta$-normalized variables are given by
\be
 \mathbf{Y_{MEM}}=[-\frac13(1+2w^2-2c+\sqrt{3}(c-2)\bar{\Sigma}_-),-\frac{1}{\sqrt{3}}-c \bar{\Sigma}_-,-s\frac{\cb_\text{MR}}{\sqrt{3}},c\cb_\text{MR},s\sqrt{3}\bar{\Sigma}_-].
\ee
This completes the transformation of the explicit solutions in Section 4 of \cite{Lim}. These spike solutions are new.
The solutions above can also be generated by starting with the Kasner solution (\ref{Kasner_reparam})
above and applying the transformations in the Subsection \ref{ad} successively.

\subsubsection{Properties of the spike}

Both electromagnetic spike solutions have non-trivial electric and magnetic fields, and the names `EME' and `MEM' indicate what Rosen transitions occur at large $x$ for $|w| > 1$.
Note that the spatial dependence of the different variables lies, as in the vacuum case, in $c$ and $s$, which depends on $x$ in a spiky way. 
The electromagnetic spike solutions have the same radius as the vacuum spike solution.
For more discussions on $c$ and $s$, and on the radius of the spike, see Sections 4.5 and 4.6 of \cite{Lim}.

In the vacuum case there is a distinction between the false spike solution and true spike solution, with the false spike solution being a rotated Taub solution. 
There is no such distinction here because the frame rotation transformation is absent in polarized/diagonal case.
Both the EME and MEM spike solutions are two different spiky solutions.

\subsubsection{Alternative derivation}\label{ad}

The solution-generating transformations in polarized electromagnetic Gowdy spacetimes are
\begin{eqnarray}
\label{eq:rotation_EM}
 \e^{-\hat{P}/2}=\frac{\e^{-P/2}}{\chi^2+\e^{-(P+\tau)}},\quad \hat{\chi}=-\frac{\chi}{\chi^2+\e^{-(P+\tau)}},
\end{eqnarray}
and
\begin{eqnarray}
\label{eq:GE_EM}
 \hat{P}=-P,\quad \hat{\chi}_{\tau}=-\e^{P-\tau}\chi_x,\quad \hat{\chi}_x=-\e^{P+\tau}\chi_{\tau}.
\end{eqnarray}
These transformations correspond to transformations (6) and (7) in \cite{Lim}, but curiously the role of solution-generating transformation has switched.
Transformation (6) in \cite{Lim} is merely a frame rotation, but here transformation (\ref{eq:rotation_EM}) is a solution-generating transformation.
On the other hand, transformation (7) in \cite{Lim} is a solution-generating transformation, but here transformation (\ref{eq:GE_EM}) is merely
a 90-degree duality rotation for the electromagnetic field (see e.g. \cite{MS} about duality rotation) followed by a switch of the coordinates $y$ and $z$.
Transformation (\ref{eq:GE_EM}), like transformation (7) in \cite{Lim}, is also very simple when expressed in $\beta$-normalized variables:
\begin{eqnarray}\label{4}
 (\hat{\Sigma}_-, \hat{N}_{\times}, \hat{\ce}, \hat{\cb})=(-\Sigma_-, -N_{\times}, \cb, \ce).
\end{eqnarray}
Thus here we do not have `false' spikes, since both spikes represent real inhomogeneous solutions, although quite similar. 

Transformation (\ref{eq:rotation_EM}) maps the Kasner solution (\ref{Kasner_reparam}) to the electric Rosen solution (\ref{ERsol_1})--(\ref{ERsol_2}); transformation (\ref{eq:GE_EM}) then maps it to the magnetic Rosen solution 
(\ref{MRsol_1})--(\ref{MRsol_2});
(\ref{eq:rotation_EM}) then maps it to the EME spike solution (\ref{EME__1})--(\ref{EME__2}); and (\ref{eq:GE_EM}) maps it to the MEM spike solution (\ref{MEM__1})--(\ref{MEM__2}). At each step, $\lambda$ is obtained by 
quadrature.

\section{Visualization}

\begin{figure}[t!]
  \begin{center}
    \resizebox{12cm}{!}{\includegraphics{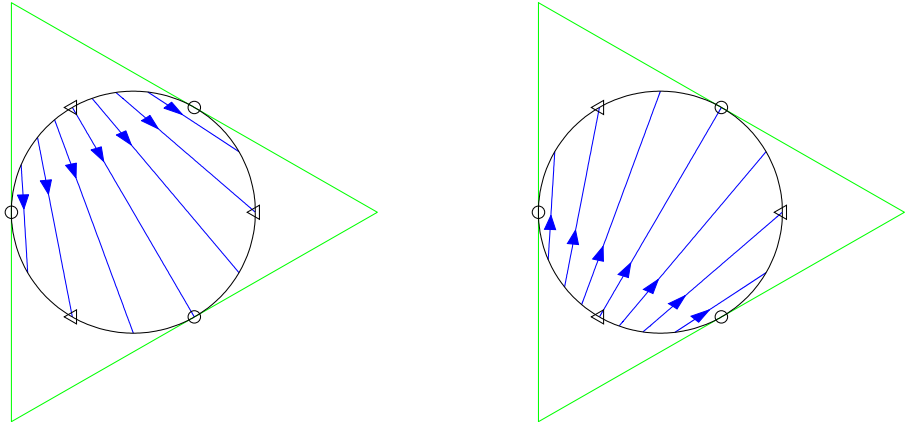}}
   \caption{The electric and magnetic Rosen orbits projected on the $(\Sp,\Sm)^H$ plane.}
    \label{fig:top}
\end{center}
\end{figure}
We have already seen that the vacuum transformations have a different interpretation in the electromagnetic case. 
Now we proceed to visualize the dynamics of the different solutions and we compare the dynamics of the
solutions found here corresponding to polarized Gowdy spacetimes with a electromagnetic field with the ones found in \cite{Lim}
corresponding to Gowdy spacetimes. For this purpose we use Hubble-normalized variables~\cite{WH,UEWE}, denoted with a superscript $H$.
For the spatially homogeneous background dynamics, it is best to use the 
Hubble-normalized variables which are related to the $\beta$-normalized ones via
\be
\label{Hubble-norm}
	(\Sigma_+,\Sigma_-,N_{\times},\ce,\cb)^H = \frac{1}{1-\Sp}(\Sigma_+,\Sigma_-,N_{\times},\ce,\cb),
\ee
and satisfy
\be
	\Sph {}^2+\Smh {}^2+N_{\times}^H {}^2+\frac13 (\ce^H {}^2+ \cb^H {}^2)=1.
\ee
The Hubble-normalized energy density $\Omega$ of the electromagnetic field is given by
\be
	\Omega = \frac13 (\ce^H {}^2+ \cb^H {}^2),
\ee
while the Hubble-normalized spatial curvature $\Omega_k$ is given by
\be
	\Omega_k = N_{\times}^H {}^2.
\ee

\begin{figure}[t!]
  \begin{center}
    \resizebox{12cm}{!}{\includegraphics{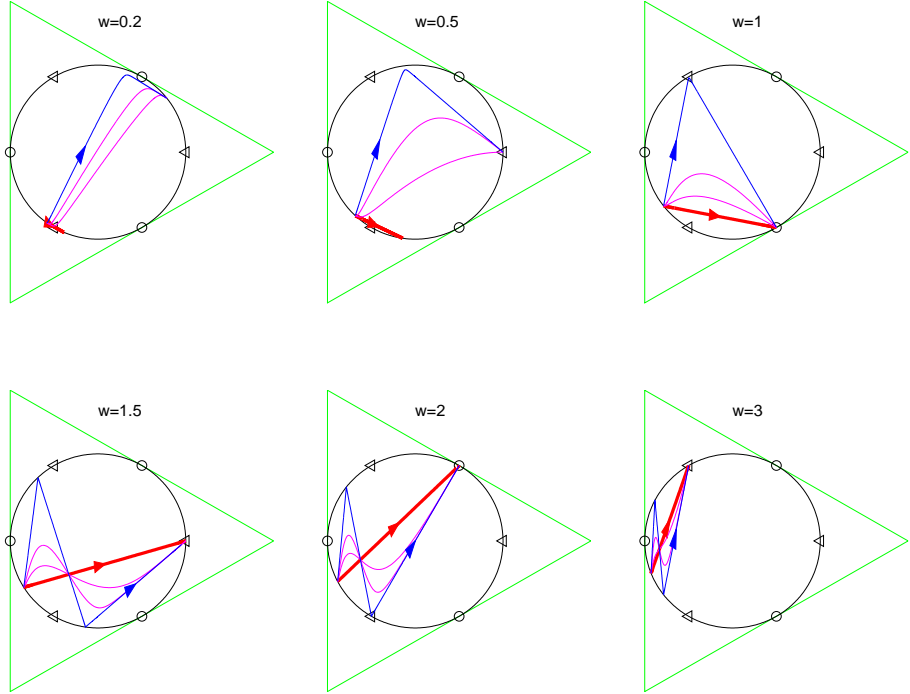}}
    \caption{Orbits of the MEM spike solution projected on the $(\Sp,\Sm)^H$ 
plane. Orbits are colored red along the spike worldline $x=0$, blue along 
$x=1000$, and magenta along small values of $x$. $w=0.2$, $0.5$, $1$, $1.5$, 
$2$, $3$ respectively.}
   \label{fig:spike_orbits}
\end{center}
\end{figure}
 The dynamics of the electric and magnetic Rosen solutions can be described by their orbits in the state space. When projected on the $(\Sigma_+,\Sigma_-)^H$ plane, these orbits form straight lines emanating 
from one corner of a triangle superscribing the Kasner circle (see Figure~\ref{fig:top}).
The two transition sets combine to describe the dynamics during an electromagnetic equivalence of a Kasner era, which consists of long Kasner epochs (described by the Kasner equilibrium points), punctuated by brief periods of transitions. Either the electric or the magnetic component becomes significant during these periods of transitions. Compare with Figure 5 of \cite{Lim}. The coincidence of the projected Rosen orbits with the projected Taub orbits compels one to compare the Rosen solutions with the Taub solution. Consider the Taub solution (24)--(26) of \cite{Lim} with a particular value for $w$ (call it $w_\text{T}$) and the corresponding magnetic Rosen solution whose orbit starts and ends at the same Kasner points as this Taub orbit. Then one finds that the $w$-parameter for the magnetic Rosen solution takes the value 
$w_\text{MR}=w_\text{T}/2$. Consequently the rate of change for the magnetic Rosen transition from one Kasner point to the next is only one half of that for the Taub transition. 

\begin{figure}[t!]
  \begin{center}
    \resizebox{12cm}{!}{\includegraphics{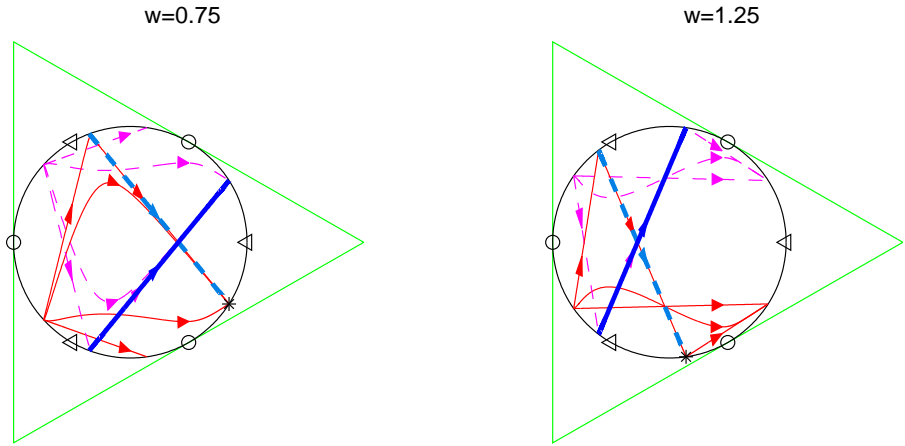}}
    \caption{Two families of orbits with $w=0.75$ and $1.25$. 
The Kasner seed is indicated by a black *, the electric Rosen orbit in dashed light 
blue, the magnetic Rosen orbit in dark blue, orbits of the EME spike solution in 
dashed magenta, and orbits of the MEM spike solution in red.}
    \label{fig:EM_family}
\end{center}
\end{figure}

The implication of this is that in more general models where both modes are present, gravitationally-driven Taub transitions would dominate electromagnetically-driven Rosen transitions towards the singularity, 
and that the Hubble-normalized energy density of the electromagnetic field would tend to zero towards the singularity, with the caveat that this occurs almost everywhere, except possibly at some `spiky' worldlines where the Taub 
mode has a local zero. Along these spiky worldlines, whether (vacuum) spike transitions would dominate Rosen transitions is unknown. 
In Figure~\ref{fig:spike_orbits} we visualize the orbits of the MEM spike solution (\ref{MEM__1})--(\ref{MEM__3}) along various worldlines $x=\text{const}$, projected on the $(\Sigma_+,\Sigma_-)^H$ plane.
The projected orbits of the EME spike solution (\ref{EME__1})--(\ref{EME__3}) differ only in the sign of $\Sm$.
Along worldlines far away from the spike worldline, the orbits approximate the electric and magnetic Rosen orbits. Along the spike worldline, the projected orbit is a straight line. For the case $|w| \geq 1$, all these orbits end at the same Kasner points. For the case $0 < |w| < 1$, the orbit along the spike worldline ends at a different Kasner point from all others.
Compare with Figure 6 of \cite{Lim}. In Figure~\ref{fig:EM_family} we visualize two families of the solutions (one with $w=0.75$ and the other with $w=1.25$). Note that the Kasner solution plotted here uses the parametrization in 
(\ref{Kasner_reparam}).
Compare with Figure 9 of \cite{Lim}.

\begin{figure}[t!] 
  \begin{center}
    \resizebox{12cm}{!}{\includegraphics{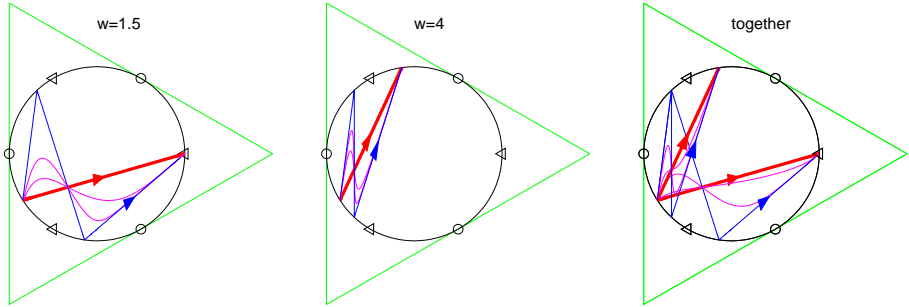}}
    \caption{Comparison of orbits for the MEM spike solution with $w_\text{MEM}=1.5$ and orbits for the vacuum spike solution with $w_\text{S}=4$. Both sets of orbits start at the same Kasner point but end at two different 
Kasner points.}
    \label{fig:EM_vec_comparison}
\end{center}
\end{figure}
We now compare the orbits of MEM spike solution with the orbits of vacuum spike solution. Call the corresponding parameters $w_\text{MEM}$ and $w_\text{S}$. We set $w_\text{S} = 2w_\text{MEM}+1$ so that both solutions start at 
the same Kasner point. But the solutions will not end at the same Kasner points. See Figure~\ref{fig:EM_vec_comparison}.

\section{Discussion and outlook}
In this paper we have presented new solutions to the Einstein-Maxwell solutions with polarized Gowdy symmetry which generalize the known magnetic and electric Rosen solutions which
are spatially homogeneous to the inhomogeneous case. These solutions represent spikes in the electromagnetic field as well as the gravitational field, and are building blocks of oscillatory behavior. In contrast to the
vacuum case there are no false and true electromagnetic spikes, however the solutions are now linked via a duality rotation. From the analysis one can see that it is the gravitational solutions
which dominate the electromagnetic solutions. Therefore we conjecture that the (vacuum) gravitational spike solution plays a larger role than the electromagnetic spike solutions in the oscillatory regime.
The analysis represents also additional support to the analysis carried out in \cite{WIB}, where an inhomogeneous generalization
of Bianchi VI$_0$ with a pure magnetic field was considered. In fact, the latter solution can be seen as coming from an electromagnetic field which does not come from a vector
potential, cf. \cite{RG}. In \cite{RG} an inhomogeneous generalization of Bianchi VII$_0$ was also presented. Both these inhomogeneous generalizations are called of local
or twisted Gowdy symmetry. It is of interest to investigate further these cases as a pre-step to the analysis of the Einstein-Maxwell system with full Gowdy symmetry which
is considerably more complicated. It remains unclear whether pure magnetic spike solutions exist.
In \cite{ColeyLim} numerical and analytical evidence was presented that (gravitational) spikes can generate matter perturbations and it was argued that this
phenomenon might explain the formation of structure in the early Universe. Electromagnetic spikes should have a similar effect on matter as well.
It remains to be seen how gravitational spikes interact with electromagnetic spikes, and whether such interactions amplify or suppress the electromagnetic field.

 \section*{Appendix}

The equations in this section are taken from from \cite{nungesser} where $-\lambda$ instead of $\lambda$ was used. 
Note also that we have put $\omega=0$ and we have used a different sign convention for $\epsilon^{\alpha\beta\gamma\delta}$.

The basic quantity in electromagnetism is the \textit{electromagnetic field tensor} $F_{\alpha\beta}$ which is antisymmetric. 
The \textit{Maxwell equations} in absence of charged matter are
\begin{eqnarray*}
\nabla^{\alpha}F_{\alpha\beta}&=&0\\
\nabla_{\alpha}F_{\beta\gamma}+\nabla_{\gamma}F_{\alpha\beta}+\nabla_{\beta}F_{\gamma\alpha}&=&0.
\end{eqnarray*}
The second Maxwell equation can be solved by introducing a four potential such that
\begin{eqnarray}\label{mx}
F_{\mu\nu} &=& \nabla_{\mu}A_{\nu}-\nabla_{\nu}A_{\mu}.
\end{eqnarray}
In the \textit{Lorentz gauge}, where by definition
\begin{eqnarray*}
 \nabla^{\alpha}A_{\alpha}&=&0
\end{eqnarray*}
 holds, the Maxwell equations in curved space time can be written as:
\begin{eqnarray}\label{qq}
 \nabla^{\alpha}\nabla_{\alpha}A_{\beta}-{{R^{\gamma}}}_{\beta}A_{\gamma}&=&0.
\end{eqnarray}
The energy-momentum tensor for an electromagnetic field is
\begin{eqnarray}\label{xs}
T_{\alpha\beta} &= &\frac{1}{4\pi} (F_{\alpha\gamma}{F_{\beta}}^{\gamma}-\frac{1}{4}g_{\alpha\beta}F_{\delta\epsilon}F^{\delta\epsilon}),
\end{eqnarray}
which is trace-free. The \textit{dual electromagnetic field tensor} is
\begin{eqnarray}\label{du}
 ^{*}F^{\gamma\delta}&=&\frac{1}{2}\epsilon^{\alpha\beta\gamma\delta}F_{\alpha\beta}.
\end{eqnarray}
where
\begin{eqnarray*}
 \epsilon^{\alpha\beta\gamma\delta}&=&(-\det g)^{-\frac{1}{2}}\eta^{\alpha\beta\gamma\delta},
\end{eqnarray*}
where $\det g$ is the determinant of the matrix $g_{\alpha\beta}$. Let $n_{\alpha}$ be a unit future-pointing vector orthogonal to a spacelike hypersurface. 
Then we can define the electric and the magnetic fields as follows:
\begin{eqnarray}\label{p1}
E^{\alpha}&=&F^{\alpha\beta}n_{\beta},\\\label{p2}
B^{\alpha}&=&^{*}F^{\alpha\beta}n_{\beta}.
\end{eqnarray}

The symmetry assumptions imply that vector potential has only the following components:
\begin{eqnarray*}
A_{3}=&\chi(\tau,x).
\end{eqnarray*}
$A_2$ is periodic of period $2\pi$ with respect to $x$. Using (\ref{mx}) one can compute that the electromagnetic field tensor has the following non-trivial components:
\begin{eqnarray*}
F_{03}& = &\chi_{\tau},\\
F_{13}& = &\chi_x.
\end{eqnarray*}
The dual electromagnetic field tensor has according to (\ref{du}) the following non-trivial components:
\begin{eqnarray*}
^{*}F^{02}&=&-\chi_x e^{\frac{3\tau-\lambda}{2}},\\
^{*}F^{12}&=&\chi_{\tau}e^{\frac{3\tau-\lambda}{2}}.
\end{eqnarray*}
Choosing 
\begin{eqnarray*}
 n_{\alpha}=(\sqrt{-g_{00}},0,0,0)
\end{eqnarray*}
as the unit future-pointing vector we compute the non-vanishing components of the electric and the magnetic field with (\ref{p1}) and (\ref{p2}):
\begin{eqnarray*}
  E^{3}&=&\chi_{\tau}e^{\frac{-\lambda+7\tau+4P}{4}},\\
 B^{2}&=&\chi_x e^{\frac{-\lambda+3\tau}{4}}.
\end{eqnarray*}
$\ce$ and $\cb$ are $\beta$-normalized orthonormal frame components of the electric and magnetic fields.
\begin{eqnarray*}
&& \ce=\frac{ e^3{}_3 E^3}{\beta}=2\chi_{\tau}e^{\frac12(P+\tau)}\\
&& \cb=\frac{ e^2{}_2 B^2}{\beta}=-2\chi_xe^{\frac12(P-\tau)},
\end{eqnarray*}
where $e^3{}_3 = e^{(-P-\tau)/2}$ and $e^2{}_2 = e^{(P-\tau)/2}$ are the frame coefficients.

\section*{Acknowledgments}
We thank Alan A. Coley, Alan D. Rendall and Claes Uggla for helpful comments and suggestions. This work was initiated when both authors were still at the 
Max-Planck-Institute for Gravitational Physics, the first author being funded through the project SFB 647 of the German Research Foundation.  E.N. is grateful to the G\"{o}ran Gustafsson Foundation for Research in Natural Sciences and Medicine and the Irish Research Council for their financial support.


\begin{thebibliography}{10}
\bibitem{BKL3}
E.~M. Lifshitz and I.~M. Khalatnikov.
\newblock {Investigation in relativistic cosmology}.
\newblock {\em Adv. Phys.}, 12:185--249, 1963.
\bibitem{BKL1}
V.~A. Belinskii, I.~M. Khalatnikov, and E.~M. Lifshitz.
\newblock {Oscillatory approach to a singular point in the relativistic
  cosmology}.
\newblock {\em Adv. Phys.}, 19:525--573, 1970.

\bibitem{BKL2}
V.~A. Belinskii, I.~M. Khalatnikov, and E.~M. Lifshitz.
\newblock {A general solution of the Einstein equations with a time
  singularity}.
\newblock {\em Adv. Phys.}, 31:639--667, 1982.

\bibitem{BergerMoncrief}
B.~K. Berger and V.~Moncrief.
\newblock {Numerical investigation of cosmological singularities}.
\newblock {\em Phys. Rev. D}, 48:4676, 1993.

\bibitem{GM}
B.~Grubi\v{s}i\'{c} and V.~Moncrief.
\newblock {Asymptotic behavior of the $T^{3} \times R$ Gowdy space-times.}
\newblock {\em Phys. Rev. D.}, 47:2371--82, 1993.

\bibitem{Bergerliving}
B.~K. Berger.
\newblock {Numerical Approaches to Spacetime Singularities}.
\newblock {\em Living Rev. Relativity}, 1:7. URL (cited on 7/15/2013), 2005.

\bibitem{RX}
A.~D. Rendall and M.~Weaver.
\newblock {Manufacture of Gowdy spacetimes with spikes}.
\newblock {\em Class. Quant. Grav.}, 18:2959--2976, 2001.

\bibitem{KR}
S.~Kichenassamy and A.~D. Rendall.
\newblock {Analytic description of singularities in Gowdy spacetimes}.
\newblock {\em {Class. Quant. Grav.}}, 15:1339--55, 1998.

\bibitem{RD}
A.~D. Rendall.
\newblock {Fuchsian analysis of singularities in Gowdy spacetimes beyond
  analyticity}.
\newblock {\em Class. Quant. Grav.}, 17:3305--3316, 2000.

\bibitem{Lim}
W.~C. Lim.
\newblock {New explicit spike solutions-non-local component of the generalized
  Mixmaster attractor}.
\newblock {\em Class. Quantum Grav.}, 25:045014, 2008.

\bibitem{Lim2}
W.~C. Lim, L.~Andersson, D.~Garfinkle, and F.~Pretorius.
\newblock {Spikes in the Mixmaster regime of $G_2$ cosmologies}.
\newblock {\em Phys. Rev. D}, 79:123526, 2009.


\bibitem{HUL}
J.~M. Heinzle, C.~Uggla, and W.~C. Lim.
\newblock {Spike oscillations}.
\newblock {\em Phys. Rev. D}, 86:104049, 2012.



\bibitem{ER}
E.~Nungesser and A.~D. Rendall.
\newblock {Strong cosmic censorship for solutions of the Einstein-Maxwell field
  equations with polarized Gowdy symmetry}.
\newblock {\em Class. Quant. Grav.}, 26:105019, 2009.


\bibitem{RA}
A.~D. Rendall.
\newblock {\em {Partial differential equations in general relativity}}.
\newblock Oxford University Press, Oxford, 2008.

\bibitem{nungesser}
E.~Nungesser.
\newblock {\em {Strong cosmic censorship in polarized $T^3$-Gowdy symmetric
  spacetimes with a Maxwell field}}.
\newblock Diploma thesis. Free University, Berlin, 2008.

\bibitem{VUW}
H.~van Elst, C.~Uggla, and J.~Wainwright.
\newblock {Dynamical systems approach to $G_2$ cosmology}.
\newblock {\em Class. Quant. Grav.}, 19:51--82, 2002.

\bibitem{CA}
M.~Carmeli, C.~Charach, and S.~Malin.
\newblock {Survey of cosmological models with gravitational, scalar and
  electromagnetic waves}.
\newblock {\em Physics Reports}, 76, 2:79--156, 1981.

\bibitem{Rosen62}
G.~Rosen.
\newblock {Symmetries of the Einstein-Maxwell Equations}.
\newblock {\em J. Math. Phys.}, 3:313--318, 1962.

\bibitem{Li2}
S.~Liebscher, A.~D. Rendall, and S.~B. Tchapnda.
\newblock {Oscillatory singularities in Bianchi models with magnetic fields}.
\newblock {\em arXiv:1207.2655 [gr-qc]}, 2012.

\bibitem{MS}
C.~W. Misner, K.~S. Thorne, and J.~A. Wheeler.
\newblock {\em {Gravitation}}.
\newblock Freeman, 1973.

\bibitem{WH}
J.~Wainwright and L.~Hsu.
\newblock {A dynamical systems approach to Bianchi cosmologies: orthogonal
  models of class A}.
\newblock {\em Class. Quant. Grav.}, 6:1409--1431, 1989.

\bibitem{UEWE}
C.~Uggla, H.~van Elst, J.~Wainwright, and G.~F.~R. Ellis.
\newblock {The past attractor in inhomogeneous cosmology}.
\newblock {\em Phys. Rev. D}, 68:103502, 2003.

\bibitem{WIB}
M.~Weaver, J.~Isenberg, and B.~K. Berger.
\newblock {Mixmaster Behavior in Inhomogeneous Cosmological Spacetimes}.
\newblock {\em {Phys. Rev. Lett.}}, 80:2984--2987, 1998.

\bibitem{RG}
A.~D. Rendall.
\newblock {Dynamics of solutions of the Einstein equations with twisted Gowdy
  symmetry}.
\newblock {\em J. Geom. Phys.}, 62:569--577, 2012.

\bibitem{ColeyLim}
A.~A. Coley and W.~C. Lim.
\newblock {Generating matter inhomogeneities in general relativity}.
\newblock {\em Phys. Rev. Lett.}, 108:191101, 2012.

\end{thebibliography}
\end{document}